\pdfoutput=1

\documentclass[prd,twocolumn,aps,amssymb,amsmath,tightenlines,floatfix]{revtex4}
\usepackage{graphicx,epsfig}

\newcommand{\cP}{\ensuremath{\mathcal{P}}}
\newcommand{\cL}{\ensuremath{\mathcal{L}}}
\newcommand{\cD}{\ensuremath{\mathcal{D}}}
\newcommand{\cT}{\ensuremath{\mathcal{T}}}
\newcommand{\cC}{\ensuremath{\mathcal{C}}}
\newcommand{\cPT}{\ensuremath{\mathcal{PT}}}
\newcommand{\half}{\mbox{$\textstyle{\frac{1}{2}}$}}
\newcommand{\fourth}{\mbox{$\textstyle{\frac{1}{4}}$}}
\newcommand{\eighth}{\mbox{$\textstyle{\frac{1}{8}}$}}
\newcommand{\threehalf}{\mbox{$\textstyle{\frac{3}{2}}$}}
\newcommand{\vep}{\varepsilon}

\begin{document}
\title{$\cPT$-symmetric quantum field theory in $D$ dimensions}

\author{Carl M.~Bender$^{a,c}$}\email{cmb@wustl.edu}
\author{Nima Hassanpour$^a$}\email{nimahassanpourghady@wustl.edu}
\author{S.~P.~Klevansky$^b$}\email{spk@physik.uni-heidelberg.de}
\author{Sarben Sarkar$^c$}\email{sarben.sarkar@kcl.ac.uk}

\affiliation{$^a$Department of Physics, Washington University, St.~Louis,
Missouri 63130, USA\\
$^b$Institut f\"ur Theoretische Physik, Universit\"at Heidelberg,
69120 Heidelberg, Germany\\
$^c$Department of Physics, King's College London, London WC2R 2LS, UK}

\begin{abstract}
$\cPT$-symmetric quantum mechanics began with a study of the Hamiltonian $H=p^2+
x^2(ix)^\vep$. A surprising feature of this non-Hermitian Hamiltonian is that
its eigenvalues are discrete, real, and positive when $\vep\geq0$. This paper
examines the corresponding quantum-field-theoretic Hamiltonian $H=\half(\nabla
\phi)^2+\half\phi^2(i\phi)^\vep$ in $D$-dimensional spacetime, where $\phi$ is a
pseudoscalar field. It is shown how to calculate the Green's functions as series
in powers of $\vep$ directly from the Euclidean partition function. Exact finite
expressions for the vacuum energy density, all of the connected $n$-point
Green's functions, and the renormalized mass to order $\vep$ are derived for
$0\leq D<2$. For $D\geq2$ the one-point Green's function and the renormalized
mass are divergent, but perturbative renormalization can be performed. The
remarkable spectral properties of $\cPT$-symmetric quantum mechanics appear to
persist in $\cPT$-symmetric quantum field theory.
\end{abstract}
\maketitle

\section{Introduction\label{s1}}
The study of $\cPT$-symmetric quantum theory may be traced back to a series of
papers that proposed a new perturbative approach to scalar quantum field theory.
Instead of a conventional perturbation expansion in powers of a coupling
constant, it was proposed that a parameter $\delta$ that measures the
nonlinearity of the theory could be used as a perturbation parameter
\cite{r1,r2}. Thus, to solve a $g\phi^4$ field theory we studied a $g\phi^2(
\phi^2)^\delta$ theory and treated the parameter $\delta$ as small. The
procedure was to obtain a perturbation expansion in powers of $\delta$ and then
to set $\delta=1$ to obtain the results for the $g\phi^4$ theory. Detailed
investigation showed that this perturbative calculation is numerically accurate
and does not require the coupling constant $g$ to be small \cite{r1,r2}. An
important feature of this approach was that $\phi^2$ and not $\phi$ had to be
raised to the power $\delta$ in order to avoid raising a negative quantity to a
noninteger power, thereby generating complex numbers as an artifact of the
procedure.

This $\delta$ expansion was also used to solve nonlinear classical differential
equations of physics \cite{r3}: the Thomas-Fermi equation (nuclear charge
density) $y''(x)=[y(x)]^{3/2}/\sqrt{x}$ becomes $y''(x)=y(x)[y(x)/x]^\delta$;
the Lane-Emdon equation (stellar structure) $y''(x)+2y'(x)/x+[y(x)]^n=0$ becomes
$y''(x)+2y'(x)/x+[y(x)]^{1+\delta}$; the Blasius equation (fluid dynamics) $y'''
(x)+y''(x)y(x)=0$ becomes $y'''(x)+y''(x)[y(x)]^\delta=0$; the Korteweg-de Vries
equation (nonlinear waves) $u_t+uu_x+u_{xxx}=0$ becomes $u_t+u^\delta u_x+u_{xxx
}=0$. In each case the quantity raised to the power $\delta$ is positive and
when $\delta=0$ the equation becomes linear. Also, these $\delta$ expansions
have a nonzero radius of convergence and are numerically accurate.

$\cPT$-symmetric quantum mechanics began with the surprising discovery that
spurious complex numbers do not appear if the quantity raised to the power
$\delta$ is $\cPT$ symmetric (invariant under combined space and time
reflection) \cite{r4,r5}. This fact is highly nontrivial and was totally
unexpected. Indeed, the eigenvalues of the non-Hermitian $\cPT$-symmetric
Hamiltonian 
\begin{equation}
H=p^2+x^2(ix)^\vep\quad(\vep\geq0)
\label{e1}
\end{equation}
are entirely real, positive, and discrete when $\vep\geq0$ because $ix$ is
$\cPT$ invariant. A proof that the spectrum is real when $\vep>0$ was given by
Dorey, Dunning, and Tateo \cite{r6,r7}. Numerous $\cPT$-symmetric model
Hamiltonians have been studied at a theoretical level \cite{r8} and many
laboratory experiments have been performed on $\cPT$-symmetric physical systems
\cite{r9,r10,r11,r12,r13,r14,r15,r16,r17,r18,r19,r20,r21}.

The purpose of this paper is to introduce powerful new tools and techniques that
can be used to investigate $\cPT$-symmetric quantum field theories. We
illustrate these tools by studying the quantum-field-theoretic analog of
(\ref{e1}) whose $D$-dimensional Euclidean-space Lagrangian density is 
\begin{equation}
\cL=\half(\nabla\phi)^2+\half\phi^2(i\phi)^\vep\quad(\vep\geq0),
\label{e2}
\end{equation}
where $\phi$ is a pseudoscalar field so that $\cL$ is $\cPT$ invariant. We treat
$\vep$ as small and show how to calculate the vacuum energy density $E_0$, the
connected $n$-point Green's functions $G_n$, and the renormalized mass $M_{\rm
R}$ as series in powers of $\vep$. In this paper we assume that $0\leq D<2$ to
avoid the appearance of renormalization infinities and then we comment briefly
on the perturbative renormalization procedure for the case $D\geq2$.

To first order in $\vep$ ($\vep<<1$), the unusual Lagrangian density $\cL$ in
(\ref{e1}) has a {\it logarithmic} self-interaction term:
\begin{equation}
\cL=\half(\nabla\phi)^2+\half\phi^2+\half\vep\phi^2\log(i\phi)+{\rm O}(\vep^2).
\label{e3}
\end{equation}
For a quantum field theory having a complex logarithmic interaction term it is
not obvious whether one can find Feynman rules for performing perturbative
diagramatic calculations. We will show how to construct such Feynman rules. We
begin by replacing the complex logarithm with a real logarithm and we do so in
such a way as to preserve $\cPT$ symmetry; to wit, we define
$$\log(i\phi)\equiv\half i\pi+\log(\phi)\quad({\rm if}~\phi>0)$$
and we define
$$\log(i\phi)\equiv-\half i\pi+\log(-\phi)\quad({\rm if}~\phi<0).$$
Combining these two equations, we make the replacement
\begin{equation}
\log(i\phi)=\half i\pi\,|\phi|/\phi+\half\log(\phi^2).
\label{e4}
\end{equation}
Note that in (\ref{e4}) the imaginary part is odd in $\phi$ and the real part
is even in $\phi$. Thus, (\ref{e4}) enforces the $\cPT$ symmetry because the
pseudoscalar field $\phi$ changes sign under parity $\cP$ and $i$ changes sign
under time reversal $\cT$. [To derive (\ref{e4}) we must assume that $\phi$ is
real. The reality of $\phi$ is explained in Sec.~\ref{s2}.]

Graphical techniques were developed in Ref.~\cite{r1} to handle real logarithmic
interaction terms. These techniques are generalizations of the replica trick
\cite{r22}, which has been used in the study of spin glasses. The idea of the
replica trick is that a logarithmic term $\log A$ can be reformulated as the
limit $\log A=\lim_{N\to0}\textstyle{\frac{1}{N}}\big(A^N-1\big)$, or
equivalently and slightly more simply as the limit
\begin{equation}
\log A=\lim_{N\to0}\textstyle{\frac{d}{dN}}A^N.
\label{e5}
\end{equation}
One then regards $N$ as an integer and identifies $A^N$ as an $N$-point vertex
in a graphical expansion. Of course, this procedure is not rigorous because it
requires taking the {\it continuous} limit $N\to0$. The validity of this
approach has not been proved, but when it is possible to compare with exactly
known results in low-dimensional theories, the replica trick gives the correct
answer. In this paper we verify our field-theoretic results by comparing them
with the exact answers for $D=0$ (where the functional integral becomes an
ordinary integral) and for $D=1$ (quantum mechanics).

The graphical calculations in this paper are done in coordinate space. Once the
vertices have been identified, all that one needs is the free propagator in
$D$-dimensional Euclidean space $\Delta(x-y)$, which satisfies the differential
equation
\begin{equation}
\big(-\nabla_x^2+1\big)\Delta(x-y)=\delta^{(D)}(x-y).
\label{e6}
\end{equation}
Taking the Fourier transform of this equation gives the amplitude for the free
propagator of a particle of mass $1$ in momentum space:
$$\tilde\Delta(p)=1/(p^2+1).$$
The $D$-dimensional inverse Fourier transform of this expression then gives the
$D$-dimensional coordinate-space propagator in terms of an associated Bessel
function:
\begin{equation}
\Delta(x_1-x_2)=(2\pi)^{-\frac{D}{2}}|x_1-x_2|^{1-\frac{D}{2}}{\rm K}_{1-
\frac{D}{2}}(|x_1-x_2|).
\label{e7}
\end{equation}

If we let $x_1\to x_2$, we obtain the amplitude $\Delta(0)$ for a {\it self
loop}, which is the amplitude for a line to originate from and return to the
same point:
\begin{equation}
\Delta(0)=(4\pi)^{-D/2}\Gamma(1-D/2).
\label{e8}
\end{equation}
This expression is finite and nonsingular for $0\leq D<2$.

This paper is organized as follows. We calculate the ground-state energy density
$E_0$ in Sec.~\ref{s2}, the one-point Green's function $G_1$ in Sec.~\ref{s3},
the two-point Green's function $G_2$ and the renormalized mass $M_{\rm R}$ in
Sec.~\ref{s4}, and the general connected $n$-point Green's function $G_n$ in
Sec.~\ref{s5}, all to first order in $\vep$. These quantities are finite when $0
\leq D<2$, but the three quantities $E_0$, $G_1$, and $M_{\rm R}$ diverge when
$D\geq2$ so it is necessary to introduce a renormalization procedure. In
Sec.~\ref{s6} we discuss the issues of renormalization. We show that a
redefinition of the energy scale, an additive shift in the field, and a mass
counterterm eliminate these infinities. In this section we also discuss our
future calculational objectives, namely, calculating the Green's functions to
higher order in $\vep$.

\section{First-order calculation of the ground-state energy density\label{s2}}
If we expand the partition function
\begin{equation}
Z(\vep)=\int\!\cD\phi\,e^{-\int\! d^D\!x\,\cL}
\label{e9}
\end{equation}
for the Lagrangian density $\cL$ to first order in $\vep$ and use
(\ref{e4}), the functional integral (\ref{e9}) becomes
\begin{eqnarray}
Z(\vep) &=& \int\!\cD\phi\,e^{-\int\! d^D\!x\,\cL_0}
\bigg(1-\frac{\vep}{4}\int\! d^Dy\,\big\{i\pi\phi(y)|\phi(y)\nonumber\\
&&\qquad+\phi^2(y)\log\big[\phi^2(y)\big]\big\}
+{\rm O}(\vep^2)\bigg),
\label{e10}
\end{eqnarray}
where the free Lagrangian density
\begin{equation}
\cL_0=\half(\nabla\phi)^2+\half\phi^2
\label{e11}
\end{equation}
is obtained by setting $\vep=0$ in $\cL$. Note that the imaginary part of the
functional integrand in (\ref{e10}) is odd in $\phi$, so $Z(\vep)$ is {\it
real}.

We emphasize that the functional integration in (\ref{e10}) is performed along
the {\it real}-$\phi$ axis and not in the complex-$\phi$ domain. This justifies
the use of (\ref{e4}). We are not concerned here with complex functional
integration paths that terminate in complex Stokes sectors because the
functional integral (\ref{e10}) converges term-by-term in powers of $\vep$. In
$\cPT$-symmetric quantum mechanics the boundary conditions on the Schr\"odinger
equation associated with the Hamiltonian (\ref{e1}) [see (\ref{a2})] are imposed
in complex Stokes sectors \cite{r5}. However, in the context of quantum field
theory it would be hopelessly unwieldy to consider {\it functional} Stokes
sectors. This is why we treat $\vep$ as small. This paper is concerned with
calculating the coefficients in the $\vep$ series and we do not consider here
the mathematical issues involved with the summation of such a series for large
values of $\vep$.

In general, a partition function is the exponential of the ground-state energy
density $E_0$ multiplied by the volume of spacetime $V$: $Z=e^{-E_0V}$. Thus,
the {\it shift} in the ground-state energy density $\Delta E$ to order $\vep$ is
given by
$$\Delta E=\frac{\vep}{4Z(0)V}\int\!\cD\phi\,e^{-\int d^D\!x\,\cL_0}
\!\int\!d^D\!y\,\phi^2(y)\log\big[\phi^2(y)\big].$$
Hence, from (\ref{e5}) we obtain
$$\Delta E=\lim_{N\to1}\frac{\vep}{4Z(0)V}\,\frac{d}{dN}
\int\!\cD\phi\,e^{-\int d^D\!x\,\cL_0}\!\int\!d^D y\,\phi^{2N}(y).$$

This expression has a graphical interpretation as the product of $N$ self loops
from the spacetime point $y$ back to $y$. There are exactly $(2N-1)!!$ ways to
construct these self loops, so the expression for $\Delta E$ simplifies to 
$$\Delta E=\lim_{N\to1}\frac{\vep}{4V} \,\frac{d}{dN}\int\!d^D\!y\,[\Delta(0)]^N
(2N-1)!!.$$
Next, we note that the integral $\int d^Dy$ is the volume of spacetime $V$, so
this formula simplifies further:
$$\Delta E=\lim_{N\to1}\frac{\vep}{4}\,\frac{d}{dN}[\Delta(0)]^N(2N-1)!!.$$
Finally, we use the duplication formula for the gamma function \cite{r23}
to write
$$(2N-1)!!=2^N\Gamma\big(N+\half\big)/\sqrt{\pi}$$
and then take the derivative with respect to $N$ to get
\begin{equation}
\Delta E=\fourth\vep\Delta(0)\left\{\log[2\Delta(0)]+
\Gamma'\big(\threehalf\big)\big/\Gamma\big(\threehalf\big)\right\},
\label{e12}
\end{equation}
where $\Gamma'\big(\threehalf\big)\big/\Gamma\big(\threehalf\big)=2-\gamma-2
\log2$. The result in (\ref{e12}) may be verified for the special cases of
$D=0$ and $D=1$.

{\it Special case} $D=0$: For $D=0$ the normalized partition function becomes an
ordinary integral, which we can expand to first order in $\vep$:
\begin{eqnarray}
Z&=& \frac{1}{Z_0}\int_{-\infty}^\infty\!d\phi\,e^{-\half\phi^2(i\phi)^\vep}
\nonumber\\
&\sim& \frac{1}{Z_0}\int_{-\infty}^\infty\!d\phi\,e^{-\half\phi^2}\big[1-\half
\vep\phi^2\log(i\phi)\big],
\label{e13}
\end{eqnarray}
where $Z_0=\sqrt{2\pi}$. If we take the negative logarithm of this result, we
obtain the first-order shift in the ground-state energy density
$$\Delta E=\frac{\vep}{2\sqrt{2\pi}}\int_{-\infty}^\infty\!d\phi\,
e^{-\half\phi^2}\phi^2\log(i\phi).$$
We then integrate separately from $-\infty$ to $0$ and from $0$ to $\infty$ and
combine the two integrals to obtain a single real integral that we evaluate
as follows:
\begin{eqnarray}
\Delta E&=&\frac{\vep}{\sqrt{2\pi}}\int_0^\infty\!d\phi\,e^{-\half\phi^2}\phi^2
\log\phi\nonumber\\
&=&\textstyle{\frac{\vep}{4}}\big[\Gamma'\big(\threehalf\big)
\big/\Gamma\big(\threehalf\big)+\log2\big].
\label{e14}
\end{eqnarray}
Taking $D=0$ in (\ref{e8}) gives $\Delta(0)=1$, so (\ref{e12}) reduces exactly
to (\ref{e14}) in zero-dimensional spacetime.

{\it Special case} $D=1$: In quantum mechanics, $\Delta E$ to leading order
in $\vep$ is the expectation value of the interaction Hamiltonian $H_I=\half\vep
x^2\log(ix)$ [see (\ref{e3})] in the unperturbed ground-state eigenfunction
$\psi_0(x)=\exp\big(-\half x^2\big)$:
\begin{eqnarray}
\Delta E&=&\frac{\vep}{2}\int_{-\infty}^\infty\! dx\,e^{-x^2}x^2\log(ix)\Big/
\!\int_{-\infty}^\infty \!dx\,e^{-x^2}\nonumber\\
&=&\eighth\vep\Gamma'\big(\threehalf\big)\big/\Gamma\big(\threehalf\big).
\label{e15}
\end{eqnarray}
Taking $D=1$ in (\ref{e8}) gives $\Delta(0)=\half$, so the general result in
(\ref{e12}) reduces exactly to (\ref{e15}) in one-dimensional spacetime.

Note that in $\cPT$-symmetric quantum mechanics the calculation of expectation
values requires the $\cC$ operator \cite{r8}. However, the $\cC$ operator is not
needed for any of the calculations in this paper because a functional integral 
involves {\it vacuum} expectation values. The vacuum state is an eigenstate of
the $\cC$ operator with eigenvalue $1$, $\cC|0\rangle=|0\rangle$, so all
reference to $\cC$ disappears. This simplification was first pointed out in
Ref.~\cite{r24}.

\section{First-order calculation of the one-point Green's function\label{s3}}
The one-point Green's function $G_1$ is {\it nonperturbative} in character but
it can be calculated by following the approach used above to calculate $\Delta
E$. Keeping terms that do not vanish under $\phi\to-\phi$, we evaluate directly
the functional-integral representation
\begin{eqnarray}
G_1(a)&=&-\frac{\vep}{4Z(0)}\int\!\cD\phi\,\phi(a)e^{-\int d^D\!x\,\cL_0}
\nonumber\\
&&\quad\times\int\!d^D\!y\,i\pi\phi(y)|\phi(y)|,
\label{e16}
\end{eqnarray}
where $\cL_0$ is the free Euclidean Lagrangian in (\ref{e11}).

We use the integral identity
\begin{equation}
\phi|\phi|=\frac{2}{\pi}\phi^2\int_0^\infty\frac{dt}{t}\sin(t\phi)
\label{e17}
\end{equation}
to replace $\phi(y)|\phi(y)|$ in the functional integral (\ref{e16}) and then we
replace $\sin(t\phi)$ by its Taylor series:
\begin{equation}
\sin(t\phi)=\sum_{n=0}^\infty\frac{(-1)^nt^{2n+1}}{(2n+1)!}\phi^{2n+1}.
\label{e18}
\end{equation}
This converts (\ref{e16}) to the product of an infinite sum in $n$, a
one-dimensional integral in $t$, a $D$-dimensional integral in $y$, and a
functional integral in $\phi$:
\begin{eqnarray}
G_1(a)&=&-\frac{i\vep}{2}\int_{t=0}^\infty dt\,\sum_{n=0}^\infty\frac{(-1)^n
t^{2n}}{(2n+1)!} \int d^D\!y\nonumber\\
&&\!\!\!\!\!\!\!\!\!\!\!\!\times\frac{1}{Z(0)}\int\cD\phi\,e^{-\int
d^D\!x\,\cL_0}\phi(a)[\phi(y)]^{2n+3}.
\label{e19}
\end{eqnarray}

The sum and multiple integrals in (\ref{e19}) may appear to be difficult, but
like the calculation of the ground-state energy density, the functional integral
in the second line of (\ref{e19}) also has a graphical interpretation; it is
merely the product of the free propagator $\Delta(a-y)$ representing a line from
$a$ to $y$ multiplied by $n+1$ self loops from $y$ to $y$, and this product is
accompanied by the combinatorial factor $(2n+3)!!$. Thus, the second line in
(\ref{e19}) reduces to $(2n+3)!!\Delta(y-a)\Delta^{n+1}(0)$.

This result simplifies further because, as we can see from (\ref{e6}), the
$D$-dimensional integral is trivial: $\int d^Dy\Delta(y-a)=1$. This establishes
the translation invariance of $G_1$. The rest is straightforward:
\begin{eqnarray}
G_1&=&-\frac{i\vep}{2}\int_{t=0}^\infty dt\,\sum_{n=0}^\infty
\frac{(-1)^n t^{2n}\Delta^{n+1} (0)(2n+3)!!}{(2n+1)!}\nonumber\\
&=&-\frac{i\vep}{2}\int_{t=0}^\infty dt\,\Delta(0)\big[3-\Delta(0)t^2\big]
e^{-\half \Delta(0)t^2}
\nonumber\\
&=&-i\vep\sqrt{\pi\Delta(0)/2}.
\label{e20}
\end{eqnarray}
This expression for $G_1$ is exact to order $\vep$.

Observe that the expression for  $G_1$ is a {\it negative imaginary} number.  
This is precisely what we would expect based on previous studies of classical
$\cPT$-symmetric systems. The classical trajectories in complex coordinate space
of a particle described by the Hamiltonian (\ref{e1}) are left-right symmetric
but they lie mostly in the lower-half plane \cite{r8}. Thus, the average value
of the classical orbits is a negative-imaginary number.

{\it Special case} $D=0$: The one-point Green's function in $D=0$ is given by
the second line in (\ref{e13}) with an additional extra factor of $\phi$:
$$G_1\sim\frac{1}{Z(0)}\int_{-\infty}^\infty d\phi\,\phi e^{-\half\phi^2}
\big[1-\half\vep\phi^2\log(i\phi)\big],$$
where $Z(0)=\sqrt{2\pi}$. The integration over the first term in the square
brackets vanishes by oddness. We evaluate the contribution of the second term by
integrating first from $-\infty$ to $0$ and then from $0$ to $\infty$.
Combining these two integrals, we obtain
$$G_1= -i\vep\sqrt{\pi/2}.$$
This result is in exact agreement with the general result in (\ref{e20})
because
$\Delta(0)=1$ when $D=0$.

{\it Special case} $D=1$: When $D=1$, the expression for $G_1$ in (\ref{e20})
reduces to
\begin{equation}
G_1=-\half i\vep\sqrt{\pi}.
\label{e21}
\end{equation}
In the Appendix we derive $G_1$ in quantum mechanics and verify (\ref{e21}).

\section{First-order calculation of the two-point Green's function\label{s4}}
To obtain the {\it connected} two-point Green's function $G_2(a,b)$, one must
subtract $G_1^2$ from the vacuum expectation value of $\phi(a)\phi(b)$. However,
we have seen that $G_1$ is of order $\vep$. Therefore, to first order in $\vep$
we need only evaluate $Z(\vep)$ in (\ref{e10}) with $\phi(a)\phi(b)$ inserted
after $\cD\phi$ and then divide this integral by $Z(\vep)$. We may neglect the
imaginary terms in these integrals because they are odd under $\phi\to-\phi$.
Thus, the expression that we must evaluate for $G_2(a,b)$ is
$$\frac{\int\cD\phi\,\phi(a)\phi(b)e^{-\int d^D\!x\,\cL_0}\left\{1-\frac{\vep}
{4}\int d^D\!y\,\phi^2(y)\log\big[\phi^2(y)\big]\right\}}{\int\cD\phi\,e^{-\int
d^D\!x\,\cL_0}\left\{1-\frac{\vep}{4}\int d^D\!y\,\phi^2(y)\log\big[\phi^2(y)
\big]\right\}}.$$
We expand this expression to first order in $\vep$ as a sum of three functional
integrals:
\begin{equation}
G_2(a,b)=A+B+C,
\label{e22}
\end{equation}
where
\begin{eqnarray}
\!\!\!\!\!\!\!\!A&=&\frac{1}{Z(0)}\int\!\cD\phi\,\phi(a)\phi(b)e^{-\int d^D\!x\,
\cL_0},\nonumber\\
\!\!\!\!\!\!\!\!B&=& -\frac{\vep}{4Z(0)}\int\!\cD\phi\,\phi(a)\phi(b)e^{-\int
d^D\!x\,\cL_0}\nonumber\\
&&\!\!\!\!\!\!\!\!\!\!\!\!\times\int\!d^D\!y\,\phi^2(y)\log\big[\phi^2(y)
\big],\nonumber\\
\!\!\!\!\!\!\!\!C&=& \frac{\vep}{4Z(0)}\int\!\cD\phi\,\phi(a)\phi(b)e^{-\int
d^D\!  x\,\cL_0}\nonumber\\
&&\!\!\!\!\!\!\!\!\!\!\!\!\times\frac{1}{Z(0)}\int\!\cD\phi\,e^{-\int d^D
\!x\,\cL_0}\!\int\!d^D\!y\,\phi^2(y)\log\big[\phi^2(y)\big].
\label{e23}
\end{eqnarray}

We must now evaluate the three contributions $A$, $B$, and $C$. The functional
integral $A$ in (\ref{e23}) is simply the free propagator $\Delta(a-b)$. This
result verifies that if we set $\vep=0$ in the Lagrangian (\ref{e2}), we obtain
a free field theory; the two-point Green's function for such a field theory is
$\Delta(a-b)$.

The double functional integral $C$ presents a complication. The first line of
$C$ is proportional to $A$ and evaluates to $\fourth\vep\Delta(a-b)$. However,
as we showed in the calculation of the ground-state energy, the next two
lines of $C$ evaluate to $\frac{4}{\vep}V\Delta E$, where $\Delta E$ is the
first-order shift in the ground-state energy density (\ref{e12}) and $V$ is the
volume of Euclidean spacetime. Thus, $C=\Delta(a-b)V\Delta E$, and this
quantity is {\it divergent} because $V$ is infinite.

We resolve this divergence problem by calculating $B$. Using (\ref{e5}) we
express the $B$ integral as
\begin{eqnarray}
B&=& -\frac{\vep}{4Z(0)}\lim_{N\to1}\frac{d}{dN}\int\cD\phi\,e^{-\int d^D\!x\,
\cL_0} \nonumber\\
&&\qquad\times\int d^D\!y\,\phi(a)\phi(b)\phi^{2N}(y).
\label{e24}
\end{eqnarray}
This functional integral requires that we connect in a pairwise fashion the set
of $2N+2$ points consisting of $a$, $b$, and the $2N$ points $y$ with the free
propagator $\Delta$ in (\ref{e7}). 

There are two cases to consider. In the first case $a$ is connected to $b$ and
the remaining $2N$ points at $y$ are connected in pairs. Note that this
reproduces the result above for $C$ except with the opposite sign. Thus, the
volume divergences exactly cancel.

In the second case $a$ is not connected to $b$. Instead, $a$ connects to a point
$y$ (there are $2N$ ways to do this) and $b$ connects to one of the remaining
$2N-1$ points $y$ (there are $2N-1$ ways to do this). The rest of the $2N-2$
points $y$ are joined in pairs [there are $(2N-3)!!$ ways to do this]. The
amplitude for this case is
$$-\frac{\vep}{2}\int d^D\!y\,\Delta(a-y)\Delta(y-b)\lim_{N\to1}\frac{d}{dN}N
(2N-1)!!\Delta^{N-1}(0),$$
which simplifies to $-\vep K\int d^D\!y\,\Delta(a-y)\Delta(y-b)$, where
\begin{eqnarray}
K&=&\half+\half\Gamma'\big(\threehalf\big)/\Gamma\big(\threehalf\big)
+\half\log[2\Delta(0)]\nonumber\\
&=&\threehalf-\half\gamma+\half\log\big[\half\Delta(0)\big].
\label{e25}
\end{eqnarray}

Thus, our final result for the coordinate-space two-point Green's function to
order $\vep$ is
\begin{equation}
G_2(a-b)=\Delta(a-b)-\vep K\int d^D\!y\,\Delta(a-y)\Delta(y-b).
\label{e26}
\end{equation}
In momentum space this is
\begin{equation}
\tilde{G}_2(p)=\frac{1}{p^2+1}-\vep\frac{K}{(p^2+1)^2}+{\rm O}\big(\vep^2\big).
\label{e27}
\end{equation}

From (\ref{e27}) we construct the $(0,1)$ Pad\'e approximant, which is just the
geometric sum of a chain of bubbles:
\begin{equation}
\tilde{G}_2(p)=\frac{1}{p^2+1+\vep K+{\rm O}\big(\vep^2\big)}.
\label{e28}
\end{equation}
We then read off the square of the renormalized mass to first order in $\vep$:
\begin{equation}
M_{\rm R}^2=1+K\vep+{\rm O}\big(\vep^2\big).
\label{e29}
\end{equation}

{\it Special case} $D=0$: To verify (\ref{e28}) and (\ref{e29}) for the case
$D=0$ we evaluate the ordinary one-dimensional integrals in
\begin{eqnarray}
G_2&=&\frac{\int dx\,x^2e^{-x^2(ix)^\vep/2}}{\int dx\,e^{-
x^2(ix)^\vep/2}}\nonumber\\
&=&\frac{\int_0^\infty dx\,x^2e^{-x^2/2} \left[1-\frac{\vep}{4}
x^2\log(x^2)\right]} {\int_0^\infty dx\,e^{-x^2/2}\left[1-\frac{\vep}{4}
x^2\log(x^2)\right]}.\nonumber
\end{eqnarray}
These integrals are not difficult and to order $\vep$ we get
\begin{equation}
G_2=\frac{1}{1+\vep\big(\threehalf -\half\gamma-\half\log2\big)}.
\label{e30}
\end{equation}
This result agrees exactly with that in (\ref{e28}) for $D=0$.

{\it Special case} $D=1$: We verify (\ref{e29}) for the case $D=1$ by
calculating the energy level of the {\it first} excited state of the
Hamiltonian $H$ in (\ref{e1}). When $\vep=0$, $H$ becomes the
harmonic-oscillator Hamiltonian, so the first excited state eigenfunction is
$xe^{-x^2/2}$ and the associated eigenvalue is $\threehalf$. We solve the
time-independent Schr\"odinger equation $H\psi=E\psi$ perturbatively by
substituting
\begin{eqnarray}
\psi(x)&=&xe^{-x^2/2}+\vep\psi_1(x)+{\rm O}\big(\vep^2\big),\nonumber\\
E&=&\threehalf+\vep E_1+{\rm O}\big(\vep^2\big),
\label{e31}
\end{eqnarray}
and collecting powers of $\vep$. To first order in $\vep$ the function
$\psi_1(x)$ satisfies
\begin{eqnarray}
&&-\half\psi_1''(x)+\half x^2\psi_1(x)-\threehalf\psi_1(x)\nonumber\\
&& \qquad=\left[-\half x^2\log(ix)+E_1\right]xe^{-x^2/2}.\nonumber
\end{eqnarray}

To solve this equation we use the technique of reduction of order and let
$\psi_1(x)=xe^{-x^2/2}f(x)$. The equation for $f(x)$ is then
$$xf''(x)+\left(2-2x^2\right)f(x)=-2E_1x+x^3\log(ix).$$
Multiplying this equation by the integrating factor $x\,\exp\big(-x^2\big)$
gives
$$\frac{d}{dx}\left[x^2 f'(x)e^{-x^2}\right]=\left[-2E_1x^2+x^4\log
(ix)\right]e^{-x^2}.$$
Therefore, if we integrate this equation from $-\infty$ to $\infty$, we obtain
an equation for $E_1$:
$$E_1=\frac{1}{4}\int_{-\infty}^\infty dx\,x^4\log\big(x^2\big)e^{-x^2}
\Big/\int_{-\infty}^\infty dx\,x^2e^{-x^2},$$
where we have replaced $\log(ix)$ by $\half\log\big(x^2\big)$.
These integrals are easy to evaluate and we get
$$E_1=\textstyle{\frac{3}{8}}\left(\textstyle{\frac{8}{3}}-\gamma-2\log2
\right).$$

Thus, the first excited eigenvalue of $H$ to order $\vep$ is
\begin{equation}
\threehalf+\textstyle{\frac{3}{8}}\vep\left(\textstyle{\frac{8}{3}}-\gamma
-2\log2\right).
\label{e32}
\end{equation}
We have already calculated the ground-state energy in Sec.~\ref{s2}:
\begin{equation}
\half+\textstyle{\frac{1}{8}}\vep(2-\gamma-2\log2).
\label{e33}
\end{equation}
The renormalized mass $M_{\rm R}$ is the first excitation above the ground
state so $M_{\rm R}$ is the difference of these two energies:
\begin{equation}
M_{\rm R}=1+\fourth\vep(3-\gamma-2\log2).
\label{e34}
\end{equation}
If we square this result and keep terms of order $\vep$, we get
\begin{equation}
M_{\rm R}^2=1+\vep\left(\threehalf-\half\gamma-\log2\right),
\label{e35}
\end{equation}
which exactly reproduces (\ref{e29}) for the case $D=1$.

\section{Higher-order Green's functions\label{s5}}
The connected three-point Green's function is given by the cumulant
\begin{eqnarray}
G_3(x,y,z)&=&\frac{\langle\phi(x)\phi(y)\phi(z)\rangle}{Z}
-\frac{\langle\phi(x)\phi(y)\rangle\langle\phi(z)\rangle}{Z^2}\nonumber\\
&&-\frac{\langle\phi(x)\phi(z)\rangle\langle\phi(y)\rangle}{Z^2}
-\frac{\langle\phi(y)\phi(z)\rangle\langle\phi(x)\rangle}{Z^2}\nonumber\\
&&+2\frac{\langle\phi(x)\rangle\langle\phi(y)\rangle\langle\phi(z)\rangle}{Z^3}.
\nonumber
\end{eqnarray}
However, to order $\vep$
\begin{eqnarray}
G_3(x,y,z)&=&\frac{\langle\phi(x)\phi(y)\phi(z)\rangle}{Z}-\Delta(x-y)G_1\nonumber\\
&&-\Delta(x-z)G_1-\Delta(y-z)G_1.\nonumber
\end{eqnarray}
The calculation of $G_3$ is somewhat tedious, but the procedure follows exactly
the calculation of the two-point Green's function. The final result after the
disconnected terms have canceled is
\begin{eqnarray}
G_3(x,y,z)&=&-i\frac{\vep\sqrt{\pi}}{\sqrt{2\Delta(0)}}\nonumber\\
&&\hspace{-1.0cm}\times\int\!d^D\!u\,\Delta(x-u)\Delta(y-u)\Delta(z-u).
\label{e36}
\end{eqnarray}

The connected four-point function is defined by the cumulant
\begin{eqnarray}
G_4(x,y,z,w)&=&-\frac{\langle\phi(x)\phi(y)\phi(z)\phi(w)\rangle}{Z}\nonumber\\
&&\hspace{-2.5cm}-\frac{\langle\phi(x)\phi(y)\phi(z)\rangle\langle\phi(w)
\rangle}{Z^2}-(\rm three~permutations)\nonumber\\
&&\hspace{-2.5cm}-\frac{\langle\phi(x)\phi(y)\rangle\langle\phi(z)\phi(w)
\rangle}{Z^2}-(\rm two~permutations)\nonumber\\
&&\hspace{-2.5cm}+2\frac{\langle\phi(x)\phi(y)\rangle\langle\phi(z)\rangle
\langle\phi(w)\rangle}{Z^3} +(\rm five~permutations)\nonumber\\
&&\hspace{-2.5cm}-6\frac{\langle\phi(x)\rangle\langle\phi(y)\rangle\langle\phi
(z)\rangle\langle\phi(w)\rangle}{Z^4}.
\nonumber
\end{eqnarray}
Again, calculating $G_4$ is tedious but the result is simply
\begin{eqnarray}
G_4(x,y,z,w)&=&-\frac{\vep}{\Delta(0)}\!\int\!d^D\!u\,\Delta(x-u)
\nonumber\\
&&\hspace{-1.0cm}\times\Delta(y-u)\Delta(z-u)\Delta(w-u).
\label{e37}
\end{eqnarray}

The pattern is now evident and with some effort we can calculate the connected
$n$-point Green's function to order $\vep$ and obtain a general formula valid
for all $n\neq2$:
\begin{eqnarray}
G_n(x_1,x_2,\ldots x_n)&=&-\half\vep(-i)^n\Gamma\big(\half n-1\big)
\nonumber\\
&&\hspace{-2.5cm}\times\big[\half\Delta(0)\big]^{1-n/2}
\int d^D\!u\,\prod_{k=1}^n\Delta\big(x_k-u\big).
\label{e38}
\end{eqnarray}
Thus, the connected Green's functions are all of order $\vep$ except for $G_2$
in (\ref{e26}), which is of order 1. Observe that (\ref{e38}) reduces to
(\ref{e20}), (\ref{e36}), and (\ref{e37}) for $n=1$, 3, and 4.

We emphasize that (\ref{e38}) holds for both odd and even $n$. Our calculation
of the odd-$n$ Green's functions uses the {\it first} term on the right side of
(\ref{e4}) and our calculation proceeds by introducing the integral
representation in (\ref{e17}) followed by using the Taylor series in
(\ref{e18}). On the other hand, our calculation of the even-$n$ Green's
functions uses the {\it second} term on the right side of (\ref{e4}) and the
calculation proceeds by applying the derivative identity in (\ref{e5}). It is
satisfying that these two strikingly different techniques lead to the single
universal formula in (\ref{e38}) for $G_n$.

\section{Discussion and Future Work\label{s6}}
The principal advance reported in this paper is that we have developed all of
the machinery necessary to calculate the Green's functions of a $\cPT$-symmetric
quantum field theory in (\ref{e2}) to any order in $\vep$. Thus, this paper
opens a vast area for future study and investigation; one can investigate the
masses of the theory (the poles of the Green's functions), scattering
amplitudes, critical indices, and so on. The Green's-function calculations done
in Secs.~\ref{s2}-\ref{s5} are exact to first order in $\vep$. However, the
procedures presented here immediately generalize to all orders in $\vep$.

Furthermore, as we show below, even in low orders the perturbation series in
powers of $\vep$ is highly accurate and it continues to be accurate for large
$\vep$. To illustrate, we calculate the one-point Green's function $G_1$ in $D=
0$ to {\it second} order in $\vep$. In $D=0$ this Green's function is a ratio of
two ordinary integrals:
\begin{equation}
G_1=\frac{\int_{-\infty}^\infty dx\,x\,\exp\left[-\half x^2\left(1+\vep L
+\half\vep^2L^2\right)\right]}{\int_{-\infty}^\infty dx\,\exp\left[-\half x^2
\left(1+\vep L\right)\right]},
\label{e39}
\end{equation}
where $L=\log(ix)=\half i\pi|x|/x+\half\log\left(x^2\right)$. Evaluating these
integrals is straightforward and the result is
\begin{eqnarray}
G_1&=&-i\vep\sqrt{\textstyle{\frac{\pi}{2}}}\left[1+\fourth\vep\left(\gamma-2-3
\log2\right)+{\rm O}\left(\vep^2\right)\right]\nonumber\\
&=& -i\vep\sqrt{\textstyle{\frac{\pi}{2}}}\left[1-0.8756\vep+{\rm O}\left(
\vep^2\right)\right].
\label{e40}
\end{eqnarray}

To check of the accuracy of (\ref{e40}) we calculate the one-point Green's
function for a cubic theory ($\vep=1$). We convert the expansion in (\ref{e40})
to a $[0,1]$ Pad\'e approximant,
$$G_1=-i\vep\sqrt{\textstyle{\frac{\pi}{2}}}\frac{1}{1+0.8756\vep},$$
and then set $\vep=1$ to obtain the approximate result $G_1=-0.6682i$.

The exact value of $G_1$ for the zero-dimensional cubic theory $\vep=1$ is given
by the ratio of integrals
$$G_1=\frac{\int_{-\infty}^\infty dx\,x\exp\left(-\half ix^3\right)}
{\int_{-\infty}^\infty dx\,\exp\left(-\half ix^3\right)}
=-i2^{\frac{1}{3}}\frac{\Gamma(2/3)}{\Gamma(1/3)}=-0.6369i.$$
Thus, the {\it two}-term $\vep$ expansion (\ref{e40}) has an accuracy of 5\%,
which is impressive for such a large value of $\vep$. This good result is
consistent with the results found in previous studies of the accuracy of the
$\vep$ expansion for various classical equations (see Ref.~\cite{r3}).

The third-order version of (\ref{e40}) is
\begin{eqnarray}
G_1&=&-i\vep\sqrt{\textstyle{\frac{\pi}{2}}}\left[1+\fourth\vep(\gamma-2-3\log2)
\right.\nonumber\\
&&~~+\textstyle{\frac{1}{192}}\vep^2\left(54\log^22+144\log2-36\gamma\log2
\right.\nonumber\\
&&~~\left.\left.-\pi^2+6\gamma^2-48\gamma+48\right)+
{\rm O}\left(\vep^3\right)\right]\nonumber\\
&=&-i\vep\sqrt{\textstyle{\frac{\pi}{2}}}\left[1-0.8756\vep+0.6447\vep^2+
{\rm O}\left(\vep^3\right)\right].
\nonumber
\end{eqnarray}
Converting the expansion above to a $[1,1]$ Pad\'e approximant and setting $\vep
=1$, we obtain the result that $G_1=-0.6213i$, which now differs from the exact
result by only $-2$\%. These numerical results strongly motivate us to extend
our studies of the $\vep$ expansion of $\cPT$-symmetric quantum field theories
to higher order in $\vep$. We will publish the higher-order calculations in a
future paper.

A second issue that that needs to be examined in depth is that of
renormalization. Because $\Delta(0)$ becomes singular when the dimension $D$ of
Euclidean spacetime reaches $2$, the one-point Green's function $G_1$ and the
renormalized mass $M_{\rm R}$ become singular. (To first order in $\vep$ the
higher Green's functions do not become infinite when $D=2$.) Thus, for $D\geq2$
we must undertake a perturbative renormalization procedure. 

For simplicity, in this paper we have worked entirely with dimensionless
quantities. However, to carry out a perturbative renormalization one must work
with the Lagrangian
$$\cL=\half(\nabla\phi)^2+\half\mu^2\phi^2+\half g\mu_0^2\phi^2
\big(i\mu_0^{1-D/2}\phi\big)^\vep$$
for which the dimensional parameters are explicit: $\mu$ is the unrenormalized
mass $\mu$, $\mu_0$ is a fixed parameter having dimensions of mass, and $g$ is a
dimensionless unrenormalized coupling constant. The mass renormalization
procedure consists of expressing the renormalized mass $M_{\rm R}$ in terms of
these Lagrangian parameters and absorbing the divergence that arises when $D\geq
2$ into parameter $\mu$. The coupling-constant renormalization procedure is
similar; we define the renormalized coupling constant $G_{\rm R}$ as the value
of the three-point or four-point Green's functions at particular values of the
external momentum and again absorb the divergence that arises into the
Lagrangian parameter $g$. One must then verify that all higher Green's functions
are finite when expressed in terms of $M_{\rm R}$ and $G_{\rm R}$. This program
will be carried out explicitly in a future paper.

\acknowledgments
CMB thanks the Heidelberg Graduate School of Fundamental Physics for its
hospitality.

\section{Appendix}
In this Appendix we calculate the one-point Green's function $G_1$ for the case
$D=1$ (quantum mechanics). In quantum mechanics $G_1$ is the expectation value
of the operator $x$ in the ground state $|0\rangle$. Thus, if $\psi(x)$ is the
(unnormalized) ground-state eigenfunction in coordinate space, we can express
$G_1$ as the ratio of integrals 
\begin{equation}
G_1 \equiv\frac{\langle0|x|0\rangle}{\langle0|0\rangle}
=\int dx\,x\psi^2(x)\Big/\int dx\,\psi^2(x).
\label{a1}
\end{equation}

For the quantum-mechanical Hamiltonian (\ref{e1}) the ground-state eigenfunction
obeys the time-independent Schr\"odinger equation 
\begin{equation}
-\half\psi''(x)+\half x^2(ix)^\vep\psi(x)=E\psi(x),
\label{a2}
\end{equation}
where $E$ is the ground-state energy. To first-order in $\vep$ this differential
equation becomes 
\begin{equation}
-\half \psi''(x)+\half x^2\psi(x)+\half\vep x^2\log(ix)\psi(x)=E\psi(x).
\label{a3}
\end{equation}

We solve this equation perturbatively by substituting
\begin{eqnarray}
\psi(x)&=&\psi_0(x)+\vep\psi_1(x)+{\rm O}\big(\vep^2\big),\nonumber\\
E&=&E_0+\vep E_1+{\rm O}\big(\vep^2\big),
\label{a4}
\end{eqnarray}
where $\psi_0(x)=\exp\big(-\half x^2\big)$, $E_0=\half$, and as derived in
(\ref{e15}), $E_1=\eighth\Gamma'(\threehalf)\big/\Gamma\big(\threehalf\big)$.
Collecting powers in $\vep$, we see that $\psi_0(x)$ and $\psi_1(x)$ satisfy the
differential equations
\begin{eqnarray}
-\half \psi_0''(x)+\half x^2\psi_0(x)-\half\psi_0(x)&=&0,\nonumber\\
-\half\psi_1''(x)+\half x^2\psi_1(x)-\half\psi_1(x)&&\nonumber\\
&&\!\!\!\!\!\!\!\!\!\!\!\!\!\!\!\!\!\!\!\!\!\!\!\!\!\!\!\!\!\!\!\!\!\!\!\!\!
\!\!\!\!\!\!\!\!\!\!\!\!\!\!\!\!\!\!=-\half x^2\log(ix)\psi_0(x)+E_1\psi_0(x).
\label{a5}
\end{eqnarray}

The equation for $\psi_1(x)$ is an inhomogenous version of the equation for
$\psi_0(x)$. Thus, we use reduction of order to solve the $\psi_1(x)$ equation;
we substitute
\begin{equation}
\psi_1(x)=\psi_0f(x)=e^{-\half x^2}f(x)
\label{a6}
\end{equation}
and obtain
$$f''(x)-2xf'(x)=x^2\log(ix)-2E_1.$$
We then multiply this equation by the integrating factor $\exp(-x^2)$
and integrate from $-\infty$ to $x$:
$$f'(x)e^{-x^2}=C+\int_{-\infty}^x \!ds\,\big[s^2\log(is)-2E_1\big]e^{-s^2}.$$
The integration constant $C$ vanishes because 
$$2E_1=\int_{-\infty}^\infty
\!\!ds\,s^2\log(is)e^{-s^2}\Big/\int_{-\infty}^\infty \!\!ds\,e^{-s^2}.$$
Thus, we find that
\begin{equation}
\psi_1(x)=e^{-\half x^2}\!\!\int_0^x\!\!dt\,e^{t^2}\!\!\int_{-\infty}^t
\!\!\!ds\big[s^2\log(is)-2E_1\big]e^{-s^2}.
\label{a7}
\end{equation}

We can now evaluate the integrals in (\ref{a1}) to first order in $\vep$.
This expression simplifies considerably because of the factor of $x$:
\begin{eqnarray}
G_1&=&-\frac{2\vep}{\sqrt{\pi}}\int_{-\infty}^\infty dx\,xe^{-x^2}\!\int_0^x
dt\,e^{t^2}\nonumber\\
&&\quad\times \int_t^\infty\!ds\big[s^2\log(is)-2E_1\big]e^{-s^2}.
\label{a8}
\end{eqnarray}
To evaluate this integral we integrate from $x=-\infty$ to $0$ and then from
$x=0$ to $\infty$ and combine the two integrals. The resulting integral
simplifies further because the logarithm in the integrand collapses to
$\log(-1)=i\pi$:
\begin{equation}
G_1=-2i\vep\sqrt{\pi}\!\int_0^\infty \!\!dx\,xe^{-x^2}\!\!\int_0^x
\!\!dt\,e^{t^2}\!\!\int_t^\infty \!\!ds\,s^2e^{-s^2}.
\label{a9}
\end{equation}

Evaluating this triple integral is not trivial, but by using some tricks it can
be calculated exactly and in closed form. We begin by interchanging the order of
the $s$ and $t$ integrals:
\begin{equation}
G_1=-2i\vep\sqrt{\pi}\!\int_0^\infty \!\!dx\,xe^{-x^2}\!\!\int_0^\infty
\!\!ds\,s^2e^{-s^2}\!\!\int_0^{{\rm min}(s,x)}\!\!dt\,e^{t^2}.
\label{a10}
\end{equation}
Next, we introduce polar coordinates $x=r\cos\theta$ and $s=r\sin\theta$ and
make the change of variable $t=rz$:
\begin{eqnarray}
G_1&=&-2i\vep\sqrt{\pi}\int_0^\infty\!\!dr\,r^5\!\int_0^{\pi/2}\! d\theta\,\cos
\theta\sin^2\theta\nonumber\\
&& \quad\times\int_0^{{\rm min}(\sin\theta,\cos\theta)}\!dz\,e^{r^2z^2-r^2}.
\label{a11}
\end{eqnarray}

The $r$ integral can now be done, and we obtain a sum of two double
integrals:
\begin{eqnarray}
G_1&=&-2i\vep\sqrt{\pi}\left[\int_0^{\pi/4}\!d\theta\,\cos\theta\sin^2\theta\!
\int_0^{\sin\theta}\!dz \frac{1}{(1-z^2)^3}\right.\nonumber\\
&&~~\left.+
\int_{\pi/4}^{\pi/2}\!d\theta\,\cos\theta\sin^2\theta\!\int_0^{\cos\theta}\!dz
\frac{1}{(1-z^2)^3}\right].
\label{a12}
\end{eqnarray}
These double integrals may be evaluated by using any algebraic manipulation code
such as Mathematica. The final result is
\begin{equation}
G_1=-\half i\vep\sqrt{\pi}.
\label{a13}
\end{equation}
This verifies the general result in (\ref{e20}) for the case $D=1$.

\end{document}